\documentclass[10pt]{article}

\begin{document}

\title{Light scattering test regarding the relativistic nature of heat}
\author{A. Sandoval-Villalbazo$^a$ and L.S. Garc\'{\i}a-Col\'{\i}n$^{b,\,c}$ \\
$^a$ Departamento de F\'{\i}sica y Matem\'{a}ticas, Universidad Iberoamericana \\
Lomas de Santa Fe 01210 M\'{e}xico D.F., M\'{e}xico \\
E-Mail: alfredo.sandoval@uia.mx \\
$^b$ Departamento de F\'{\i}sica, Universidad Aut\'{o}noma Metropolitana \\
M\'{e}xico D.F., 09340 M\'{e}xico \\
$^c$ El Colegio Nacional, Centro Hist\'{o}rico 06020 \\
M\'{e}xico D.F., M\'{e}xico \\
E-Mail: lgcs@xanum.uam.mx}

\maketitle

\begin{abstract}
 The dynamic structure factor of a simple relativistic fluid is
calculated. The coupling of acceleration with the heat flux
present in Eckart's version of irreversible relativistic
thermodynamics is examined using the Rayleigh-Brillouin spectrum
of the fluid. A modification of the width of the Rayleigh peak
associated to Eckart's picture of the relativistic nature of heat
is predicted and estimated.
\end{abstract}

\section{Introduction}
The interpretation of heat in the theory of relativity is a
question that has often been ignored in the literature. Besides
Tolman's exhaustive study of the problem dating from seventy years
ago \cite{Tolman} it is hardly met in a book on the subject. In
1940, C. Eckart proposed a formalism to deal with the relativistic
properties of fluids constructing an energy-matter tensor
including heat \cite{Eckart}. Since this inclusion allows an
identification of heat as mechanical energy, there is a conflict
with the standard tenets of irreversible thermodynamics.

On the other hand, the present authors introduce heat into the
formalism \cite{M1} using the same methodology as it appears in
Meixner's version of irreversible thermodynamics \cite{Groot}.
This idea has not yet met universal acceptance. In this paper we
propose an experiment using the scattering of light  by a
relativistic fluid which would throw light on the two proposals.
It is shown that the Rayleigh peak ought to be broadened by a
measurable correction if Eckart is correct. It is perhaps likely
that its features may be detected in some features of the cosmic
background radiation.
\section{Basic formalism in Eckart's thermodynamics}

The starting point in Eckart's irreversible thermodynamics is the
stress-energy tensor:
\begin{equation}
T_{\beta }^{\alpha }=\frac{n\varepsilon }{c^{2}}u^{\alpha
}u_{\beta }+ph_{\beta }^{\alpha }+\Pi _{\beta }^{\alpha
}+\frac{1}{c^{2}}q^{\alpha }u_{\beta }+\frac{1}{c^{2}}u^{\alpha
}q_{\beta }.  \label{E1}
\end{equation}
Here, $\varepsilon $ is the internal energy per particle, $u$ is
the hydrodynamic four-velocity, $p$ is the local pressure, $\Pi
_{\beta }^{\alpha }$ is the Navier tensor and $q^{\alpha }$ is the
heat four-flux. $ h_{\beta }^{\alpha }=\delta _{\beta }^{\alpha
}+\frac{1}{c^{2}}u^{\alpha }u_{\beta }$ is the spatial projector,
where $u^{\alpha }u_{\alpha }=-c^{2}$ is the speed of light. The
dissipative contributions to the stress-energy tensor in Eq.
(\ref{E1}), according to Eckart's formalism, are assumed to
satisfy the orthogonality relations:
\begin{equation}
u_{\alpha }\Pi _{\beta }^{\alpha }=u^{\beta }\Pi _{\beta }^{\alpha
}=0,\qquad u_{\alpha }\,q^{\alpha }=u^{\beta }\,q_{\beta
}=0.\label{E2}
\end{equation}
Eqs. (\ref{E2}) prevent the existence of dissipation in the time
axis. The number of particles per unit of volume $n$ satisfies a
conservation equation, so that
\begin{equation}
\dot{n}+nu_{;\alpha }^{\alpha }=\dot{n}+n\theta =0.  \label{E35}
\end{equation}
The total energy balance equation is obtained from the expression
$u^{\beta }T _{\beta ;\alpha }^{\alpha }=0$, which  as shown in
Ref. \cite{M1} reads as
\begin{equation}
n\dot{\varepsilon}\,+p\theta +q_{;\alpha }^{\alpha
}+\frac{1}{c^{2}}\dot{u} _{\alpha }q^{\alpha }+u_{;\alpha }^{\beta
}\Pi _{\beta }^{\alpha }=0. \label{E6}
\end{equation}
The entropy balance equation is obtained by means of the local
equilibrium hypothesis, $s=s(n,\varepsilon)$, so that
\begin{equation}
n\dot{s}=n\left( \frac{\partial s}{\partial n}\right)
_{\varepsilon }\dot{ \varepsilon}+n\left( \frac{\partial
s}{\partial \varepsilon }\right) _{n} \dot{n},  \label{E8}
\end{equation}
where $s$ is the entropy per particle. Using well known
thermodynamical relations and Eqs. (\ref{E35}-\ref{E6}), Eq.(
\ref{E8}) becomes after some rearrangements,
\begin{equation}
(nsu^{\alpha }+\frac{q^{\alpha }}{T})_{;\alpha }=-\frac{u_{;\alpha
}^{\beta }\Pi _{\beta }^{\alpha }}{T}-\frac{q^{\alpha }T_{,\alpha
}}{T^{2}}-\frac{1}{ c^{2}}\frac{\dot{u}_{\alpha }q^{\alpha }}{T},
\label{E10}
\end{equation}
which leads to identify the total entropy flux as $J_{[S]}^{\alpha
}=nsu^{\alpha }+\frac{q^{\alpha }}{T}$, and the local entropy
production $\sum $ as
\begin{equation}
\sum =-\frac{\theta \Pi}{T}-\frac{\sigma _{\alpha }^{\beta
}{\stackrel{o}{\Pi _{\beta }^{\alpha }}} }{T}-\frac{q^{\alpha
}}{T^{2}}(T_{,\alpha }+\frac{T\;\dot{u}_{\alpha }}{c^{2} })
.\label{E12}
\end{equation}
In Eq. (\ref{E12}) $\Pi _{\beta }^{\alpha }$ is assumed symmetric
and has been separated into its scalar and traceless components
$\Pi h_{\beta }^{\alpha }$ and $\stackrel{o}{\Pi _{\beta }^{\alpha
}}$. The traceless symmetric part of the velocity gradient is
denoted as $\sigma _{\alpha }^{\beta }$. As mentioned in Ref.
\cite{M1}, linear constitutive equations consistent with Eqs.
(\ref{E2}) read:
\begin{equation}
\Pi =-\eta _{B}\theta,   \label{C1}
\end{equation}
\begin{equation}
\stackrel{o}{\Pi ^{\alpha \beta }}=-\eta _{s}h^{\mu \alpha }h^{\nu
\beta }\sigma _{\mu \nu },  \label{C2}
\end{equation}
\begin{equation}
q^{\alpha }=-k_{th}h^{\alpha \beta }(\tau _{,\beta
}+\frac{T\;\dot{u}_{\beta }}{c^{2}}).  \label{C3}
\end{equation}
In Eqs. (\ref{C1}-\ref{C3}) the transport coefficients are $\eta
_{B}$, the bulk viscosity, $\eta _{s}$ the shear viscosity and
$k_{th}$ is the thermal conductivity. The entropy production
(\ref{E12}) and the constitutive equation (\ref{C3}) differ from
their counterparts in Meixner's scheme \cite{M1} by the presence
of the acceleration term $\frac{T\;\dot{u}_{\beta }}{c^{2}}$ whose
appearance can be traced back to the inclusion of heat in the
energy momentum tensor, a feature not present in Meixner's
formalism. The effect of this term in the RB spectrum will be the
central subject of the next section.

\section{Linearized equations and the RB sepctrum.}

As it has been repeatedly shown in the literature \cite{Berne},
the transport equations derived from Meixner's scheme arise from
the conservation laws supplemented by the constitutive equations
(\ref{C1}-\ref {C3}). If a local variable $X$ is assumed to posses
an equilibrium value $ X_{o}$ and a fluctuation around this
equilibrium value $\delta X$, then we write $X=X_{o}+\delta X$.
Assuming a fluid with vanishing hydrodynamic velocity
$u^{\alpha}=(0,0,0,c)$ , the linearized particle conservation
equation (\ref{E35}) reads
\begin{equation}
\frac{\partial }{\partial t}(\delta \rho )+\rho _{o}\delta \theta
=0. \label{L1}
\end{equation}
The linearized internal energy equation (\ref{E6}) can be written
in terms of the temperature, leading to the expression
\begin{equation}
\rho _{o}\frac{\partial }{\partial t}(\delta
T)\,+\frac{T_{o}(\gamma -1)}{ \rho _{o}}\frac{\partial }{\partial
t}(\delta \rho )-\gamma D_{th}[\nabla ^{2}(\delta
T)+\frac{T_{o}\;}{c^{2}}\frac{\partial \theta }{\partial t}]=0,
\label{L2}
\end{equation}
where $\gamma =\frac{c_{p}}{c_{\rho}}$ is the heat capacities
ratio and $D_{th}= \frac{k_{th}}{\rho _{o}C_{\rho }}$is the
thermal diffusivity. The linearized divergence of the equation of
motion reads:
\begin{equation}
\frac{\partial }{\partial t}(\delta \theta )+\frac{C_{T}^{2}}{\rho
_{o}} \nabla ^{2}(\delta \rho )-\alpha C_{T}^{2}\nabla ^{2}(\delta
T)-D_{v}\nabla ^{2}(\delta \theta )=0.  \label{L3}
\end{equation}
$C_{T}^{2}=\left( \frac{\partial p}{\partial \rho }\right) _{T}$
is the square of the isothermal speed of sound, $\alpha
=-\frac{1}{\rho o}\left( \frac{\partial \rho }{\partial T}\right)
_{p}$is the thermal expansion coefficient and $D_{v}=\frac{\eta
}{\rho _{o}}$ is the kinematic viscosity; for an ideal gas $\alpha
=\frac{1}{T_{o}}$ . Volumetric viscosity has been neglected as
well as second order contributions to the fluctuations. On the
other hand, the linearized equations that arise from Meixner's
formalism resemble equations (\ref{L1}-\ref{L3}) except for the
term $\frac{C_{T}}{c^{2}} D_{th} \gamma \omega^2 $ which appears
in the energy equation. The two sets of equations may be solved,
as shown in Ref. \cite{Berne} by going into the $(\vec{q},\omega)$
space. The procedure is the standard one, a system of three
equations for the three unknowns $ \hat{T}(\vec{q},\omega)$,
$\hat{\rho}(\vec{q},\omega)$  and $\hat{\theta} (\vec{q},\omega)$
is obtained. The corresponding determinant reads as follows:
\begin{equation}
A=\left[
\begin{array}{lll}
\omega & \rho _{o} & 0 \\
\frac{C_{T}^{2}}{\rho _{o}}q^{2} & \omega +D_{v}q^{2} &
-\frac{C_{T}^{2}}{
T_{o}}q^{2} \\
\frac{T_{o}}{\rho _{o}}(\gamma -1)\omega & -\gamma
\frac{D_{th}T_{o}}{c^{2}} \omega & \omega +\gamma D_{th}q^{2}
\end{array}
\right].  \label{s1}
\end{equation}
The matrix (\ref{s1}), has been written for Eckart's equations,
which differ only from its Meixner's counterpart in the $3-2$
component of the matrix $A$. The full dispersion relation, $\det
(A)=0$, reads:
\begin{equation}
\begin{array}{l}
\omega ^{3}+(D_{v}q^{2}+\gamma D_{th}q^{2}-\frac{C_{T}^{2}}{c^{2}}
D_{th}q^{2}\gamma )\omega ^{2}+
\\
(-C_{T}^{2}q^{2}\gamma +D_{th}D_{v}q^{4}\gamma )\omega
-C_{T}^{2}D_{th}q^{4}\gamma =0.
\end{array}
\label{D1}
\end{equation}
The only addition in the dispersion relation arising from the
inclusion of heat in the stress tensor regarding the Rayleigh peak
is $\frac{ C_{T}^{2}}{c^{2}}D_{th}q^{2}\gamma $ in the $\omega
^{2}$ term. The ordinary width of the Rayleigh peak corresponds to
the real solution of Eq. (\ref{D1} ) when $\gamma \rightarrow 1$,
namely, $\omega \simeq -D_{th}q^{2}$ .

The modification to this root is proportional to
$\frac{C_{T}^{2}}{c^{2}}$ and the new root can be fairly
approximated as:
\begin{equation}
\omega \sim -D_{th}q^{2}\left[ 1+\frac{C_{T}^{2}}{c^{2}}\right].
\label{D3}
\end{equation}
Thus, for an ideal gas, the relative change of the width in the
Rayleigh peak at temperature $T$ is proportional to the
relativistic parameter $z=\frac{kT}{mc^{2}}$. This parameter is a
natural measure of relativistic effects in systems such as the hot
electron gas present in galaxy clusters.

\section{Final remarks}

In a typical light scattering experiment \cite{Berne}, the width
of the central peak is of the order of $10^{7}Hz$. The absolute
associated change in this width, according to Eckart's version of
relativistic irreversible thermodynamics would be about $10^{3}Hz$
for a an ideal electron gas with temperature $T \sim 10^{8}K$,
with a non-vanishing thermal conductivity. This change is, in
principle, measurable and a way to verify or discard
experimentally Eckart's picture of the relativistic nature of
heat.

The authors wish to thank A.L. Garc\'{\i}a Perciante for valuable
comments and discussions. This work has been supported by CONACyT
project 41081-F and FICSAC, M\'{e}xico (PFSA).

\end{document}